\documentclass[achemso,aip,jcp,preprint,superscriptaddress,floatfix]{revtex4-1}
\usepackage[utf8]{inputenc}
\usepackage{color}
\usepackage{graphicx}
\usepackage{mathtools}
\usepackage{multirow}
\usepackage{amsmath}
\usepackage{amssymb}
\usepackage{palatino}
\usepackage{amsmath}
\usepackage{tikz}
\usetikzlibrary{shapes.geometric, arrows}

\newcommand{\ket}[1]{\left|#1\right>}

\newcommand{\kPsi}{\ket{\Psi}}						
\newcommand{\allStates}{\sigma_1,\ldots,\sigma_L}	
\newcommand{\alldim}{a_1,\ldots,a_{L-1}}                     
\newcommand{\ONvec}{\ket{\sigma_1 \ldots \sigma_L}}	
\newcommand{\ONstring}{\sigma_1 \ldots \sigma_L}		

\begin{document}
\title[]{Efficient reconstruction of CASCI-type wave functions for a DMRG state using quantum information theory and genetic algorithm}
\author{Zhen Luo}
\affiliation{
    Key Laboratory of Mesoscopic Chemistry of MOE, School of Chemistry and Chemical Engineering, Nanjing University, Nanjing 210023, China
}
\author{Yingjin Ma}
\email{yingjin.ma@sccas.cn}
\affiliation{
    Department of High Performance Computing Technology and Application Development, Computer Network Information Center, Chinese Academy of Sciences, Beijing 100190, China
}
\author{Chungen Liu}
\affiliation{
    Key Laboratory of Mesoscopic Chemistry of MOE, School of Chemistry and Chemical Engineering, Nanjing University, Nanjing 210023, China
}
\author{Haibo Ma}
\email{haibo@nju.edu.cn}
\affiliation{
    Key Laboratory of Mesoscopic Chemistry of MOE, School of Chemistry and Chemical Engineering, Nanjing University, Nanjing 210023, China
}
\date{\today}

\begin{abstract}        
We improve the methodology to construct a complete active space-configuration interaction (CAS-CI) expansion for density-matrix renormalization group (DMRG) wave function using matrix-product state representation, inspired by the sampling-reconstructed CAS [SR-CAS, Boguslawski et al, \textit{J. Chem. Phys.} \textbf{2011}, 134, 224101] algorithm. In our scheme, a genetic algorithm, in which the ”crossover” and ”mutation” process can be optimized based on quantum information theory, is employed when reconstructing the CASCI-type wave function in the Hilbert space. Test analysis results for the ground and excited state wave functions of conjugated molecules and transition metal compounds illustrate that our scheme is very efficient for searching the most important CI expansions in large active spaces.
\end{abstract}

\maketitle

\section{Introduction}
In recent years, \emph{ab initio} density matrix renormalization group (DMRG)\cite{white1992density,white1992real,mitrushenkov2001quantum,chan2002highly,legeza2003controlling,legeza2003qc,legeza2003optimizing,legeza2004quantum,chan2004algorithm,moritz2005convergence,moritz2005relativistic,rissler2006measuring,legeza2008applications,chan2009density,marti2010density,chan2011density,ma2013assessment,legeza2015advanced,chan2016matrix} method, which was originally introduced by White\cite{white1992density,white1992real} for solid state physics, emerges as a promising quantum chemical approach which can deal with active spaces (ASs) larger than those by traditional full configuration interaction (FCI)/completely active space-configuration interaction (CAS-CI) methods. In DMRG, the \textit{M} eigenvectors with largest eigenvalues of the reduced density-matrix (RDM) of sub-systems constitute a renormalized basis set for the purpose of reducing the freedom of the full configurational space. Because its computational cost scales only polynomially \cite{chan2002highly}, i.e. $O(k^{3}M^{3})+ O(k^{4}M^{2})$, where \textit{k} is the number of active orbitals. DMRG is nowadays regarded as an efficient alternative to the FCI or the CAS-CI method which has an exponential scaling. For example, a $(50e, 50o)$ AS is explored by Hachmann \textit{et al.}’s DMRG calculation\cite{hachmann2007radical} and their results firstly illustrated the polyradical nature in higher acenes; and later Mizukami et al.\cite{mizukami2012more} further extended the radical nature study to GNRs by $(84e, 84o)$ DMRG calculations, and demonstrated that the interactions among large number of $\pi$ electrons is responsible for mesoscopic size effect, which in turn lead to the polyradical nature in GNRs. Considering DMRG’s significant advantage of computational efficiency and accuracy, quantum chemists have successively developed many DMRG-based multi-configuration (MC) or multi-reference (MR) approaches, such as DMRG with self-consistent field (DMRG-SCF) \cite{zgid2008density, ghosh2008orbital, sun2017, wouters2014communication, ma2016scf,ma2015density}, DMRG with perturbation theory (DMRG-CASPT2 and DMRG-NEVPT2) \cite{kurashige2011second-order, guo2016, freitag2017multireference, phung2016cumulant}, DMRG with canonical transformation theory (DMRG-CT) \cite{yanai2010multireference, neuscamman2010strongly}, DMRG with MR configuration interaction (CI) theory \cite{saitow2013multireference, saitow2015fully}, as well as linear response theory for the density matrix renormalization group (LR-DMRG) \cite{dorando2009analytic,nakatani2014linear}. These approaches have already been implemented for researches in many fields, such as transition metal complexes \cite{freitag2015orbital}, catalytic metalloenzymes \cite{kurashige2013entangled} and aromatic excimers\cite{shirai2016computational}.

The large capacity of AS in DMRG calculation benefits the deep view of the electronic structure for a chemist. The concepts of quantum entropy/entanglement and the general quantum information theory (QIT) now are constituting an integral part in the toolbox for measuring orbital interactions and analyzing a DMRG wave function\cite{rissler2006measuring}. Comparing to the commonly used electron density (i.e. one-particle RDM or its natural occupation numbers \cite{Ziesche1999Pair}, the two-particle RDM or its cumulant in terms of the Frobenius norm \cite{Zhen2005Entanglement, Pelzer2011Strong, Luzanov2004Weyl}), the entropy/entanglement quantifies the intrinsic interaction among orbitals in the AS. For example, single orbital entropy indicates that how much an orbital can be entangled with the rest of the orbital space. In addition, the entanglement (two-orbital mutual information) determines how orbitals interact with each other, i.e., it gives the correlation between two orbitals as they are both embedded in the whole orbital space. It can illustrate the nature of calculated system in the electron-correlation level, and as such, the QIT can be an unique tool for analyzing bonding property \cite{boguslawski2013orbital, mottet2014quantum}, distinguishing static or dynamical correlation \cite{boguslawski2012entanglement, stein2016the}, choosing AS automatically \cite{stein2016automated,stein2017automated}, and dissecting chemical reactions \cite{duperrouzel2014a, stein2016character} etc. For example, Boguslawski \textit{et al.} \cite{boguslawski2013orbital} demonstrated that entanglement analysis is convenient to dissect these electron correlation effects and to provide a conceptual understanding of bond-forming and bond-breaking processes.
Szilv{\'a}si \textit{et al.} \cite{T2015Concept} also showed that species of chemical bond (e.g. covalent bond, donor-acceptor dative bond, charge-shift bond etc) and aromaticity can also be distinguished using QIT.

Nevertheless, one should notice that the DMRG wave functions are usually considered to be not directly comparable with traditional single reference (SR) or MR wave functions based on CI electronic configurations. That is because the expansion items in DMRG wave function are the so-called matrix-product states (MPSs)\cite{mcculloch2007density, schollwock2011density, nakatani2013efficient, szalay2015tensor}, which are renormalized throughout the DMRG "sweep" procedure, rather than the distinguishable electronic configurations in forms of Slater determinants (SDs). In 2007, Moritz \textit{et al.} \cite{moritz2007decomposition} rationalized a method to decompose MPS into a SD basis, however the full CI expansion for a DMRG wavefunction in a large AS with more than 20 active orbitals would be prohibitive because the number of CI SDs would be greater than $10^{12}$. Later they proposed a Monte-Carlo based sampling reconstructed CAS (SR-CAS) algorithm for \cite{boguslawski2011construction} generating the determinants. The analysis of arduengo carbene in their work \cite{boguslawski2011construction} suggests that only a comparatively small amount of SDs within the entire large AS has to be considered to construct an efficient CASCI-type wave function, and the small amount of SDs could already represent the main feature for a specific electronic state.

Within the SR-CAS framework, a predefined reference (usually the HF determinant) is used to generate the trial determinants in the AS. However, the SR determinant may not be adequate or very efficient as the reference for the molecule that owns strong multi-configuration character, such as excited states, transition metal/rare earth compounds. Herein, we propose a genetic algorithm, in which the multiple SDs can be used as the reference and the "crossover" process is employed rather than randomly Monte-Carlo process for generating new SDs. Additionally, inspired from QIT, in which the orbital interactions are quantitatively evaluated, we also introduce QIT into the "mutation" process for the purpose of generating more important excitation SDs with the explicit consideration of orbital entanglements instead of random exciting the electrons in the original SR-CAS algorithm. It can be expected that the efficiency of determinants re-construction will benefit from using the "sampling/evolutionary direction" pointed out by QIT.

The paper is scheduled as following: In Sec. \uppercase\expandafter{\romannumeral2}, we present a brief description of 1) the MPS ansatz and the CI re-construction under such ansatz, 2) the theory of orbital entanglement in quantum chemistry and 3) details about our entanglement-driving genetic algorithm (EDGA) for efficient reconstruction of CASCI-type wave functions for a DMRG state. The test analysis for typical conjugated molecules (polyacetylene, polyacene) and transition metal compounds CuCl$_{2}$ and Eu-BTBP(NO$_3$)$_3$ is presented in Sec. \uppercase\expandafter{\romannumeral3}. Finally, we draw our conclusions in Sec. \uppercase\expandafter{\romannumeral4}.

\section{Theory and Methodology}
\subsection{MPS structure and determinant reconstruction}
In a traditional CI language one can express an arbitrary electronic state $\kPsi$\ spanned by $L$\ orbitals as a linear combination of occupation number vectors $\ket{\boldsymbol{\sigma}}$, with the CI coefficients $c_{\ONstring}$\ as expansion coefficients, 
\begin{equation}\label{eq:CI_wave_function}
    \kPsi = \sum\limits_{{\boldsymbol{\sigma}}} c_{\boldsymbol{\sigma}} \ket{\boldsymbol{\sigma}} = \sum\limits_{\allStates} c_{\ONstring} \ONvec\ .
\end{equation}
The basis states $|\sigma_l\rangle$ has four possible occupation status as $\left|\uparrow\downarrow\right>, \left|\uparrow\right>, \left|\downarrow\right>, \left|0\right>$\ for the $l$-th spatial orbital. 
Turning to the MPS $ansatz$\cite{mcculloch2007density, schollwock2011density, nakatani2013efficient, szalay2015tensor}, the CI coefficients $c_{\ONstring}$\ can be encoded as a product of $m_{l-1}\times m_{l}$-dimensional matrices 
$M^{\sigma_l} = \{M^{\sigma_l}_{a_{l-1}a_l}\}$ 
\begin{align}
    \kPsi &= \sum_{\allStates} \sum_{\alldim} M^{\sigma_1}_{1 a_1} M^{\sigma_2}_{a_1 a_2} \cdots M^{\sigma_L}_{a_{L-1} 1} \ONvec = \sum_{\boldsymbol{\sigma}} M^{\sigma_1} M^{\sigma_2} \cdots M^{\sigma_L} \ket{\boldsymbol{\sigma}}\ ,\label{eq:MPS2}
\end{align}
where the first and the last matrices are $1\times m_1$-dimensional row and $m_{L-1}\times 1$-dimensional column vectors, respectively.
Collapsing the summation over the $a_l$\ indices as matrix-matrix multiplications result in the last equality. 

Moritz \textit{et al.} \cite{moritz2007decomposition} presented a method for the determination of all determinants weights during DMRG sweeps in the MPS $ansatz$: all determinants weights are saved in the very first step and, during the actual calculating process, the determinants weight will be changed together with the basis states change in every DMRG micro-iteration step because of the renormalization. In this case, one can obtain the weight of a certain determinant basis state as
\begin{equation}
    c_{\sigma_1...\sigma_L} = M^{\sigma_1}[\sigma_1]M^{\sigma2}[\sigma_2]... M^{\sigma_L}[\sigma_L]
\end{equation}
where $M$ matrices for basis transformations are obtained and kept in DMRG sweeps.

For further details, we suggest the readers to refer their original paper\cite{moritz2007decomposition}.

\subsection{Orbital Entanglement in QIT}
The orbital interaction can then be quantitatively evaluated by employing concepts from QIT.
The single-orbital entropy (or one-orbital entropy), which is the quantitative measure of entanglement provided by the von Neumann entropy, can be express as
\begin{equation}
    S^{(1)}_{i} = - \sum_{\alpha = 1}^{4} \omega_{\alpha i} \rm ln\omega_{\alpha i}
\end{equation}
where $\omega_{\alpha i}$ are the eigenvalues of the one-orbital RDM $\rho^{(1)}_{ii^{'}}$ of a given orbital $i$, 
\begin{equation}
    \rho^{(1)}_{ii^{'}} = \sum_{n} \langle n | \langle i | \Phi \rangle \langle \Phi | i^{'} \rangle | n\rangle
\end{equation}
Similarly we can obtain the two-orbital entropy
\begin{equation}
    S^{(2)}_{ij} = - \sum_{\alpha = 1}^{16} \omega_{\alpha ij} \rm ln\omega_{\alpha ij}
\end{equation}
where $\omega_{ij}$ are eigenvalues of the two-orbital RDM, and the upper limit of summation indicates 16 different occupation states of two orbitals.

According to the subadditivity property of von Neumann entropy $S$, for two certain orbitals $i$ and $j$ we can obtain
\begin{equation}
    S^{(2)}_{ij} \leqslant S^{(1)}_{i} + S^{(1)}_{j}
\label{vn_entropy}
\end{equation}
The equality in Eq.(\ref{vn_entropy}) holds only when $i$ and $j$ are not entangled. Rissler\cite{rissler2006measuring} pointed out that the difference between $S^{(1)}_{i} + S^{(1)}_{j}$ and $S^{(2)}_{ij}$ could indicate the entanglement between a pair of orbitals, and the orbital-pair mutual information is given by
\begin{equation}
    I_{ij} = \frac{1}{2}(S^{(1)}_{i} + S^{(1)}_{j} - S^{(2)}_{ij})(1 - \delta_{ij})
\end{equation}
where $\delta_{ij}$ is the Kronecker delta function to ensure that there is none mutual information for a certain orbital and itself, and the factor $1/2$ prevents interactions from being counted twice.

\subsection{Combine QIT with genetic reconstruction of determinants}
Our EDGA is introduced here as a sampling reconstruction process, aiming to find the determinants with large coefficients efficiently. The algorithm is presented as following (the flowchart is illustrated in Fig.~1):

1) Generate $N$ SDs as the initial generation.

2) Compute CI coefficients ($c_i$) for the current generation. 

3) If a certain SD has the CI coefficient larger than the predefined threshold $\eta$ (e.g. $10^{-6}$), put this SD into record.

4) If $1 - \sum_{i}^{record} c^{2}_{i} < 10^{-k}$ with the value $k$ as the pre-defined threshold is satisfied, go to Step 8; otherwise go to Step 5.

5) "Crossover" operation for generating the SDs for next generation. Randomly select two SDs with large coefficients, and generate a new SD as the combination of one's alpha spin-orbitals and the other's beta spin-orbitals. Using roulette selection method, a certain determinant $i$ has the probability $\rho ={\tilde{c}_{i}}/{\sum_{i}^{N}{\tilde{c}_{i}}}$ to be chosen. The fitness parameter $\tilde{c}_{i}$ should be carefully determined since it significantly affects the performance of the algorithm. In our implementation, we use CI coefficients $c_{i}$ of determinants treated in different ways as the measurements of fitness in different stages of the genetic algorithm to avoid early maturity. Specifically, we use $\tilde{c}_{i} = sin(|c_{i}|)$ in the first 15\% evolution steps to reduce the probabilities of those determinants with large coefficients to be selected and to increase the genetic diversity of the population; in the next 40\% steps the fitness of a certain determinant is measured by $\tilde{c}_{i} = |c_{i}|$; and we use $\tilde{c}_{i} = c_{i}^{2}$ in the rest of steps to ensure that determinants with large coefficients have obvious advantages so that the algorithm tends to converge. It is clear that in all these stages determinants with larger coefficients always have greater chance to be chosen as seeds to produce new determinants.

6) "Mutation" operation for generating the next population. Randomly change the determinants obtained from Step-5, the "Crossover" step. The occupation status of orbital $i$ and $j$ would exchange according to the probability $\rho_{ij} = {I_{ij}}/{\sum_{i}{I_{ij}}}$ with $I_{ij}$ as the mutual information. It must be noted that the mutation probability affects the randomness of the algorithm. A larger value of mutation probability makes the algorithm have a greater chance of jumping out of the local optimal solution, but also makes the genetic advantages obtained in the crossover process being undermined. Mutation probability values vary for different systems, but in general should not be too large. We can always run the algorithm for a small population size and few loops to determine a proper mutation rate.

7) Replace the current generation with the newly generated one, go back to Step 2.

8) Output the determinants in the collected record.

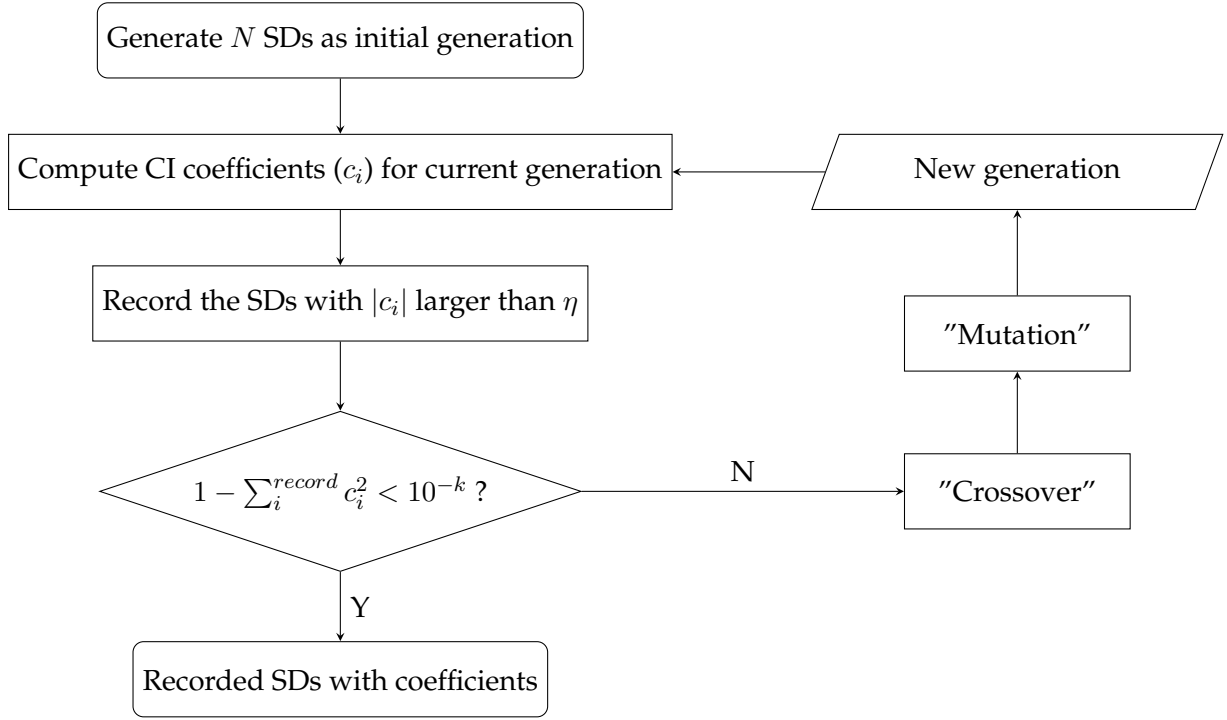
\begin{figure}[htb]	
    \label{fig_EDGA}
    \tikzstyle{startstop} = [rectangle, rounded corners, minimum width = 2cm, minimum height=1cm,text centered, draw = black]
    \tikzstyle{io} = [trapezium, trapezium left angle=70, trapezium right angle=110, minimum width=2cm, minimum height=1cm, text centered, draw=black]
    \tikzstyle{process}  = [rectangle, minimum width=3cm, minimum height=1cm, text centered, draw=black]
    \tikzstyle{decision} = [diamond, aspect = 3, text centered, draw=black]
    \tikzstyle{arrow} = [->,>=stealth]
    \begin{tikzpicture}[node distance=1cm]
    \node[startstop](start){Generate $N$ SDs as initial generation};
    \node[process, below of = start, yshift = -0.75cm](in1){Compute CI coefficients ($c_i$) for current generation};
    \node[process, below of = in1 , yshift = -0.75cm](pro1){Record the SDs with $|c_i|$ larger than $\eta$};
    \node[decision,below of = pro1, yshift = -1.5cm](dec1){$1-\sum_{i}^{record}c^{2}_{i}<10^{-k}$ ? };
    \node[process, right of = dec1, xshift = 8cm, yshift =  0cm](pro2){"Crossover"};
    \node[process, right of = pro1, xshift = 8cm, yshift = -0.40cm](pro3){"Mutation"};
    \node[io, right of = in1,  xshift = 8cm, yshift = 0cm](io1){New generation};
    \node[startstop, below of = dec1, yshift = -1.5cm](out1){Recorded SDs with coefficients};
    \coordinate (point1) at (-3cm, -6cm);
    \draw [arrow] (start) -- (in1);
    \draw [arrow] (in1) -- (pro1);
    \draw [arrow] (pro1) -- (dec1);
    \draw [arrow] (dec1) -- node [right] {Y} (out1);
    \draw [arrow] (dec1) -- node [above] {N} (pro2);
    \draw [arrow] (pro2) -- (pro3);
    \draw [arrow] (pro3) -- (io1);
    \draw [arrow] (io1) -- (in1);
    \end{tikzpicture}	
    \caption{The flowchart of the entanglement-driving genetic algorithm (EDGA) for CASCI-type wave function reconstruction.}
\end{figure} 

The key steps are Step-5 and Step-6, as the so-called Genetic Operations, in which the quantum entanglements are precisely employed via the probability. Because mutual information of orbitals acts as the selection weights in the mutation process of changing the orbital occupation status, larger value of mutation probability means greater influence of the orbital entanglements. The chosen determinants are varied by randomly choosing some pairs of spin orbitals of the same spin symmetry, and swapping the occupation status of them. The number of chosen pairs of spin orbitals are randomly determined in the searching process. This procedure could be an excitation or de-excitation process if one of the chosen spin orbitals in a pair is occupied while the other one is unoccupied, or bring no changes to the determinant if both of the two spin orbitals are occupied or unoccupied.

\section{Computational details}
\begin{figure}[htb]
    \begin{center}
        \includegraphics[width=\textwidth/2]{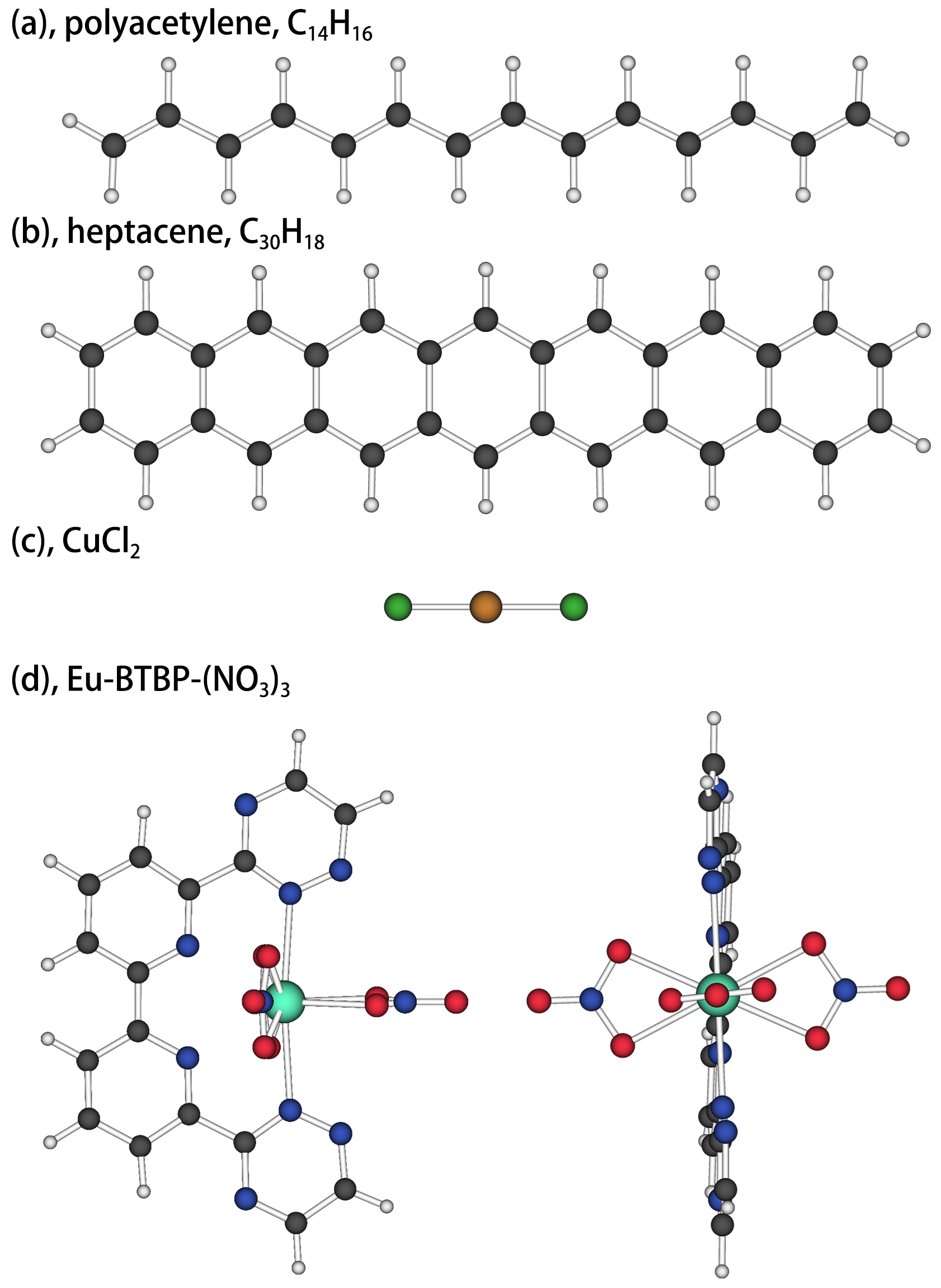}
    \end{center}
    \vspace{-0.5cm}
    \caption{Geometries of Polyacetylene (C$_{14}$H$_{16}$, a), heptacene (C$_{30}$H$_{18}$, b) and CuCl$_{2}$ (c) molecules, and Eu-BTBP(NO$_3$)$_3$ complex (d, with different orientations) that used in this paper.}
    \label{objects_geom}
\end{figure}

The CI expansion for the wavefunctions of the ground and excited states in polyacetylene (C$_{14}$H$_{16}$, Fig.~\ref{objects_geom}(a)), heptacene (C$_{30}$H$_{18}$, Fig.~\ref{objects_geom}(b)), CuCl$_{2}$ (linear, centrosymmetric geometry with the bond length $r($Cu-Cl$) = 2.154\AA{}$, Fig.~\ref{objects_geom}(c)) molecules as well as  6,6'-bis([1,2,4]-triazin-3-yl)-2,2'-bipyridine complex (Eu-BTBP-(NO$_3$)$_3$, Fig.~\ref{objects_geom}(d)) are tested in the paper.
The geometries of polyacetylene and heptacene were both optimized for $S_{0}$ state by density functional theory (DFT) method at the level of uwb97xd/6-311++g(d,p) using {\sc{Gaussian09}}\cite{g09} under their highest point group symmetry, and the geometry of Eu-BTBP-(NO$_3$)$_3$ is from Ref.\cite{narbutt2012selectivity} with the two far-end ethyls removed and the symmetry constrained to $C_{2}$ group.
The second-order Douglas-Kroll-Hess (DKH2) Hamiltonian \cite{Wolf2002TRANSGRESSING, Reiher2004Exact} in combination with ANO-RCC basis sets and a triple-$\zeta$ contraction scheme (ANO-RCC-VTZP) \cite{Widmark1990Density} was used for the CuCl$_{2}$.
For Eu-BTBP-(NO$_3$)$_3$ complex, the same DKH2 Hamiltonian and ANO-RCC with double-$\zeta$ basis sets (ANO-RCC-VDZP) were used for Eu, N, and O elements; while the ANO-RCC-MB basis sets were used for C and H elements.  
All DMRG calculations were carried out with the {\sc{QCMaquis}} DMRG software package \cite{Knecht2016New, Keller2016Spin}, which is interfaced to a development version of the quantum chemistry software package {\sc{Molcas}}\cite{Aquilante2015Molcas}. 
For the orbital basis used for DMRG calculation will be illustrated specifically for each calculation.
The DMRG-SCF calculations were also implemented using this version of {\sc{Molcas}}.

For the EDGA part, it is developed based on a local version of {\sc{QCMaquis}} DMRG software package. In addition, there are some customized setups when employing this algorithm in the work: at the beginning of the routine we generate about $N = 2000$ determinants as initialized population, which obtained by a standard Monte Carlo process with the ``most contributed determinant(s)'' based on chemical intuition acting as the initial seeds. The most contributed determinant is that in which electrons occupy those orbitals with as lower energy as possible. For instance, it is always the Hartree-Fock determinant for the ground states of closed-shell systems, and for excited states or open-shell systems it is the determinant with the fewest single-occupied orbitals and all doubly-occupied orbitals of low energy. Based on the selected determinant(s) as initial guess, we perform a few Monte Carlo steps to generate the required number of determinants as the initialized population of EDGA process. Besides, there is another way to generate initial population without Monte Carlo process. Based on the ``most contributed determinant'', we chose a small number of orbitals which closed to the highest occupied molecular orbital (HOMO) and lowest unoccupied molecular orbital (LUMO) as well as few electrons, for example $(6e, 6o)$, treat those orbitals and electrons as a small complete AS and find all the possible determinants in it. Besides, we chose another few orbitals and electrons outside of the small core AS and randomly fill the electrons into those chosen orbitals in order to generate determinants. For example, we chose 3 doubly-occupied orbitals HOMO, HOMO - 1, HOMO - 2 and 3 unoccupied orbitals LUMO, LUMO + 1 and LUMO + 2, as the $(6e, 6o)$ core AS, and another $(6e, 6o)$ outside of the core AS as the random-filling space. Ignoring the restriction of electronic state symmetry, we get hundreds of different arrangements of orbitals and electrons after traversing the core AS, and get another tens by randomly picking arrangements in the random-filling space. After that we combine the two subspaces and obtain determinants as the initialized population of our routine. In some cases the second method has a better performance.

\section{Results}
\subsection{Polyacetylene}
The conjugated molecule of polyacetylene (C$_{14}$H$_{16}$) was chosen as the first test molecule to show the performance of our EDGA searching procedure. The chosen activate space consisting of 14 valence $\pi$ orbitals and 14 valence electrons are practicable for the CASSCF calculation, so we can examine the reliability of our DMRG calculation and EDGA procedure by comparing with CASSCF results. The DMRG calculation was performed using the valence $\pi$ electrons and orbitals with the number of renormalization states $M$, we denote it as DMRG(14,14)[M]-CI.
Both canonical molecular orbitals (CMOs) and localized molecular orbitals (LMOs) were used as the base for DMRG calculations.
The CMOs was obtained by CASSCF(14,14)/6-31G(d) calculation based on the Hartree-Fock CMOs, which was implemented by using MOLPRO \cite{MOLPRO_brief}. 
The localized molecular orbitals (LMOs) was also obtained by MOLPRO employing the Pipek-Mezey method\cite{pipek1989fast} after HF-SCF calculation.

As the start, the number of renormalization states are assessed for the DMRG-CI calculation, 
in order to guarantee the weights of reconstructed SDs can exactly match these of CASSCF.
The most significant SDs from the reference CASSCF(14,14) calculations and these from DMRG(14,14)[M]-CI ($M=500,1500,5000$) are listed in Tab.~\ref{cas_14_can_gs_list}. It can be found that with increasing the number of $M$, both the coefficients of SDs and the total electronic energy become more and more close to the reference values, and the deviations can be the vanishingly small when $M = 5000$.

\begin{table}[htb]
    \label{cas_14_can_gs_list}
    \caption{Selected CI coefficients of polyacetylene for the ground state, and the difference between DMRG-EDGA CI coefficients and these of CASSCF.}
    \begin{tabular}{cccccccc}
        \hline
        \hline
        \multirow{2}*{Determinant$^{[a]}$} & \multirow{2}*{CASSCF} & \multicolumn{3}{c}{$\Delta_{\rm coeff} \times 10^{-6}$} \\ 
        \cline{3-5}
        &           & $M = 500$ & $M = 1500$ & $M = 5000$\\
        \hline
        $\rm |22222220000000\rangle$ &  0.814347 &   2227    &   121      &  0        \\
        $\rm |22222202000000\rangle$ &  0.142935 &  -1458    &  -125      &  0        \\
        $\rm |22222duud00000\rangle$ &  0.088890 &  -816     &  -66       &  0        \\
        $\rm |22222uddu00000\rangle$ &  0.088890 &  -816     &  -66       &  0        \\
        $\rm |22222020200000\rangle$ &  0.074570 &  -965     &  -76       &  0        \\
        $\rm |22222u2d000000\rangle$ &  0.072698 &   356     &   45       &  0        \\
        $\rm |22222d2u000000\rangle$ &  0.072698 &   356     &   45       &  0        \\
        $\rm |222222u0d00000\rangle$ &  0.071458 &   339     &   47       &  0        \\
        $\rm |222222d0u00000\rangle$ &  0.071458 &   339     &   47       &  0        \\
        $\rm |2222u2dd0u0000\rangle$ &  0.060828 &  -663     &  -60       &  0        \\
        $\rm |2222d2uu0d0000\rangle$ &  0.060828 &  -663     &  -60       &  0        \\
        \hline
        total energy (Hartree) & -539.5483542 & -539.5482748  & -539.5483538  & -539.5483542   \\
        \hline
        \hline  
        \multicolumn{5}{l}{$^{[a]}$ : Doubly occupied orbitals are denoted as ``2'', empty orbitals as ``0'',}   \\
        \multicolumn{5}{l}{ \ \ \  and ``u'', ``d''  are used for singly occupied orbitals by a spin-up or }   \\
        \multicolumn{5}{l}{ \ \ \  spin-down electron, respectively.}   
    \end{tabular}  
\end{table}

\begin{figure}[htb]
    \begin{center}
        \includegraphics[width=\textwidth]{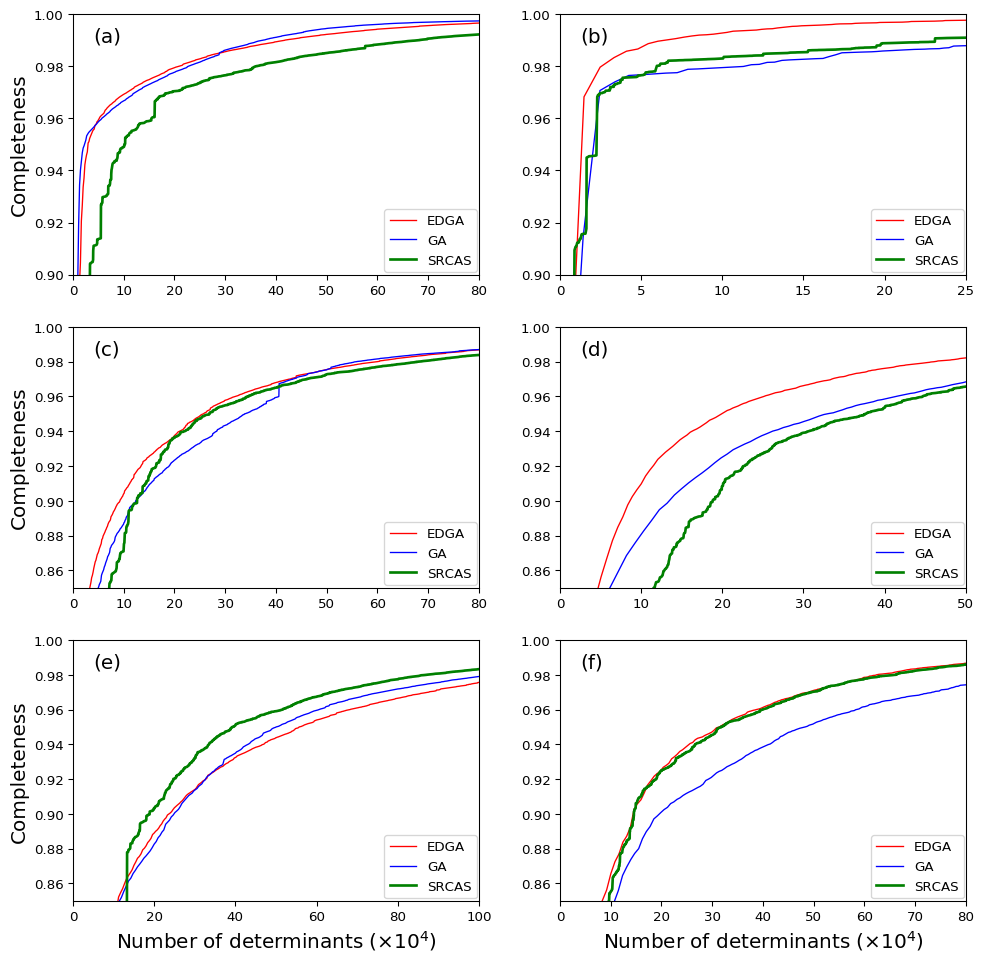}
    \end{center}
    \caption{Comparison of completeness of original GA, EDGA and SRCAS sampling schemes respecting to the number of determinants with coefficients larger than $1e^{-6}$ collected in the sampling routine. The EDGA scheme uses entanglement entropy data of canonical (left) and localized (right) orbitals from DMRG(14,14)[5000]-CI calculations on the ground state $S_{0}$ (a, b) and two excited states $S_{1}$ (c, d) and $S_{2}$ (e, f) of polyacetylene molecule.}
    \label{cas14}
\end{figure}

Basing on the converged MPS obtained from DMRG(14,14)[5000]-CI calculation, the relationship between the completeness and the number of collected determinants for different sampling schemes is investigated.
Both the original GA-based sampling scheme, our EDGA and the SRCAS are implemented. 
The results are illustrated in Fig.~\ref{cas14}(a,c,e) for canonical molecular orbitals (CMOs) and Fig.~\ref{cas14}(b,d,f) for localized molecular orbitals (LMOs), respectively.
For the former case of ground state $S_{0}$, the GA-based sampling algorithms can achieve the same completeness using around 2/3 amount of determinants comparing to the SRCAS scheme, while for excited states $S_{1}$ and $S_{2}$ the efficiency of GA almost corresponds with SRCAS. It proves the feasibility of using GA algorithm as a sampling reconstruction process. For LMOs, the GA-based sampling algorithms also have good efficiency when dealing with the $S_{0}$ and $S_{1}$ states, and the EDGA routines have noteworthy efficiency than the ones without using entanglement information.
It proves that the advantage of mutating strategy that used in GA algorithm. 
At the same time, one may notice that the entanglement doesn't have the same effect for different situations.
It implies the remaining completeness (or we can define them as the "long tail") should caused by a super huge amount and well-distributed excitations.
However, in the latter case, one can found that the GA-based sampling together with entanglements (denoted as entanglement-driving GA) show great efficiency when comparing the other algorithms. 
It implies that in localized orbital case, the ``long tail'' for the completeness can be characterized by the electronic correlation or entanglement, 
which explains the good efficiency for the EDGA algorithm.

As a short conclusion for this part, the improved GA-based searching procedure can reach the same completeness with considerable smaller amount of determinants comparing to SR-CAS algorithm in many cases.
It shows a higher space-searching efficiency, especially for the EDGA sampling scheme.

\subsection{Heptacene}
The acene consisting of adjacent benzene rings with shared bonds, is one of the another types of widely studied conjugated systems because of the controversy over its polyradical feature for long oligomers\cite{hachmann2007radical, huzak2013benchmark}. We choose the heptacene, a chain acene with 7 rings, as our research object and analyze the wave functions of singlet state S$_{0}$ with all valence orbitals and electrons.
The DMRG(30,30)[2000]-CI calculations were implemented basing on the HF CMOs from MOLPRO using $D_{2h}$ symmetry.

\begin{figure}[htb]
    \begin{center}
        \includegraphics[width=\textwidth]{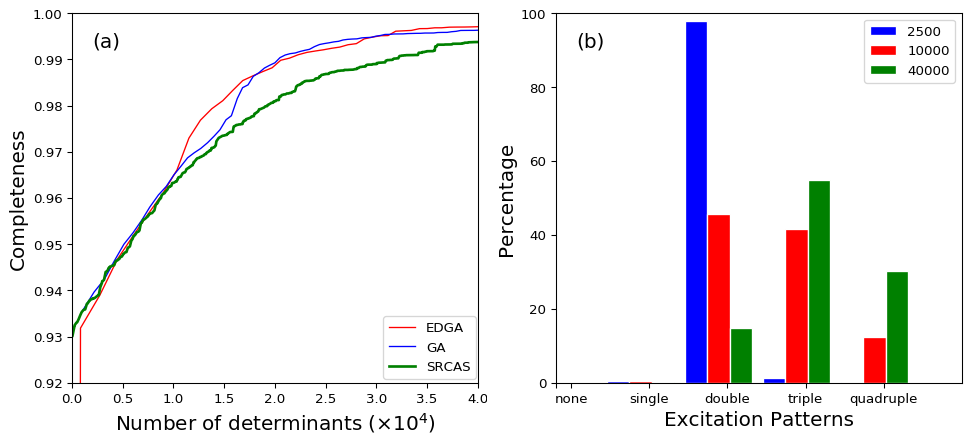}
    \end{center}
    \caption{(a) Comparison of completeness of original GA, EDGA and SRCAS schemes respecting to the number of determinants with coefficients larger than $1e^{-6}$ collected in the sampling routine. (b) The percentage of different excitation patterns among the $M = 2500, 10000, 40000$ most contributed determinants respecting to the Hartree-Fock determinant.}
    \label{cas30}
\end{figure}

The relationship between the completeness and the number of collected determinants were shown in Fig.~\ref{cas30}(a). 
It can be easily found that both the original GA and the EDGA schemes own good efficiency. The EDGA procedure reached 99.5\% completeness ($\sum_{i}c^{2}_{i} = 0.995$) by collecting about 20,000 determinants, 
while the SRCAS scheme required more than 37,000 determinants to achieve the same completeness. 
Notice that the total number of determinants within this AS is about $10^{16}$, 
it means we could explore only $1/10^{12}$ of the total configuration space but already achieve a very good approximation for the total wavefunction.
After resorting all the determinants we collected in the EDGA searching procedure by the absolute values of the coefficients from large to small, Fig.~\ref{cas30}(b) shows the percentage of various kinds of excitation patterns (number of excitations) related to different numbers of most contributed determinants. 
It can be clearly found that with more determinants being counted, the proportion of multi-excitations with small coefficients increases obviously. In detail, the percentage of quadruple excitation patterns grows from 0.2\% of 2,500 counted determinants to 12.4\% of 10,000 and 30.1\% of 40,000, despite the fact that the square sum of coefficients of all these counted determinants increases slightly from 0.9979 to 0.9982 and 0.9984.

\subsection{CuCl$_{2}$}
Next example comes to the linear, centrosymmetric transition metal compound CuCl$_{2}$. 
Because of the significant multi-configurational character of this transition metal systems, incisive analysis for the electronic structure is highly appreciated.
The active space for CuCl$_{2}$ comprises 21 electrons in 17 orbitals (Cu $3d4s3d^{'}$, and two sets of Cl $3p$ orbitals). 
The added $d$ shell, consisting of a linear combination of Cu $3d$ and $4d$ orbitals, is denoted as $3d^{'}$ that added to the Cu valence orbital space. 
The double-$d$ shell are important for a balanced description of the differential electron correlation effects in the $3d^{10}$ and $3d^{9}$ super-configurations of Cu for both ground and excited states \cite{fischer1977oscillator}. 
The state-averaged DMRG(21,17)[1000]-SCF calculation was performed with targeted five $^2a_{g}$ states of the $D_{2h}$ symmetry with equal weights.

\begin{figure}[htb]
    \begin{center}
        \includegraphics[width=\textwidth]{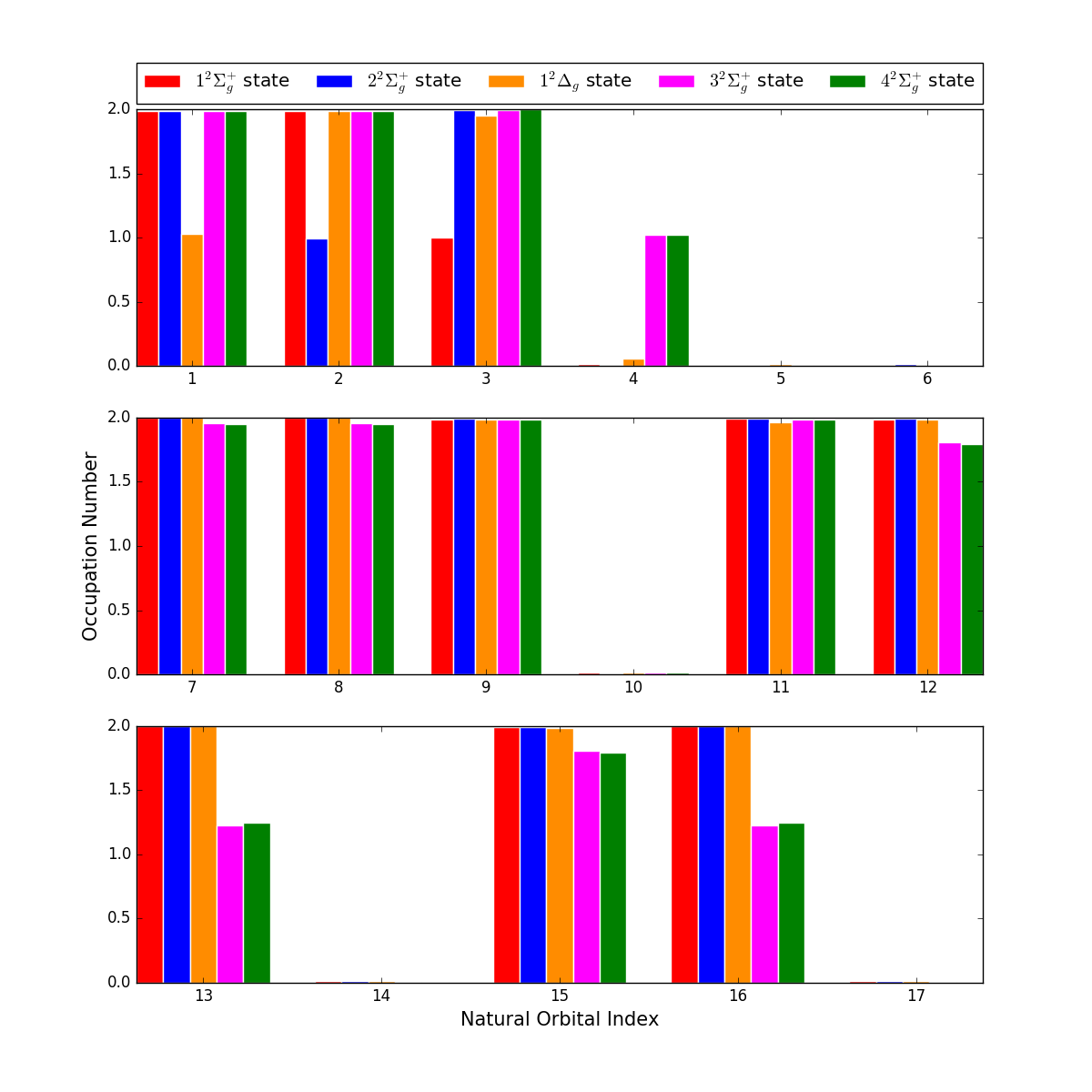}
    \end{center}
    \caption{Calculated NOONs for states $1^{2}\Sigma^{+}_{g}$, $2^{2}\Sigma^{+}_{g}$, $1^2\Delta_{g}$, $3^{2}\Sigma^{+}_{g}$, and $4^{2}\Sigma^{+}_{g}$, with orbitals ordered and grouped by irreducible representations of $D_{2h}$ symmetry.}
    \label{CuCl2_Occupation}
\end{figure}

\begin{figure}[htb]
    \begin{center}
        \includegraphics[width=\textwidth]{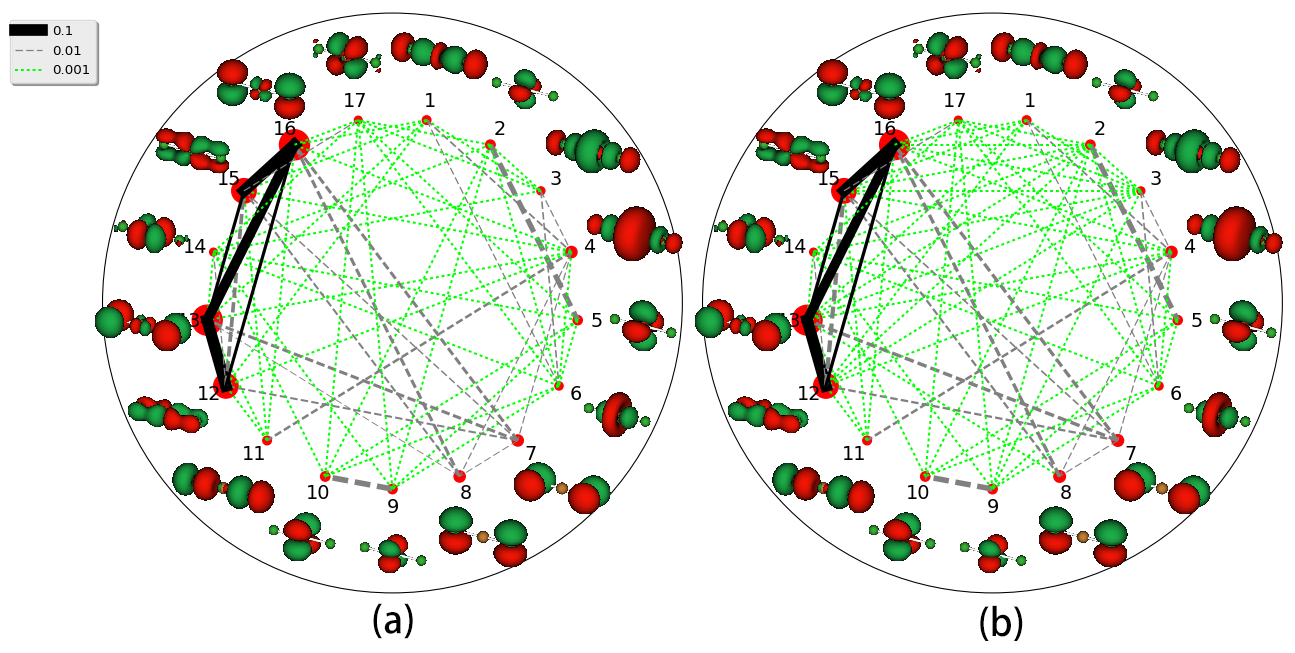}
    \end{center}
    \caption{The entanglements between canonical orbitals of states $3^{2}\Sigma^{+}_{g}$ (a) and $4^{2}\Sigma^{+}_{g}$ (b) of CuCl$_{2}$, and the related molecular orbitals. The size of the red circles indicates the magnitude of the values of single-orbital entropy $S^{(1)}_{i}$, and the color and thickness of the lines indicate mutual information $I_{ij}$.}
    \label{CuCl2_entanglement}
\end{figure}

Natural orbital occupations numbers (NOONs) of the five states are shown in Fig.~\ref{CuCl2_Occupation}.
It can be found that most states can be easily distinguished basing on the NOONs, except the $3^{2}\Sigma^{+}_{g}$ and $4^{2}\Sigma^{+}_{g}$ states.
Both of the two states have 3 singly-occupied orbitals, and NOONs are very close to each other.
Their orbital entropies and entanglements are listed in Fig.~\ref{CuCl2_entanglement}. 
However, one may notice that it still can't differentiate these two states.
In this situation, the analysis of reconstructed determinants can be helpful to understand the results.
Tab.~\ref{CuCl2_excitation_patterns} listed the important occupation determinants of these two states. 
It can be found that $3^{2}\Sigma^{+}_{g}$ and $4^{2}\Sigma^{+}_{g}$ states have the same occupation determinants with the absolute value of coefficients nearly equal. It implies that these two states are different linear combinations of the same CI SDs. Carefully distinguishing the entangled orbital pairs in Fig.~\ref{CuCl2_entanglement} and referring to the determinants in Tab.~\ref{CuCl2_excitation_patterns}, we can find that electron excitations prefer to occur in these entangled orbital pairs rather than in unentangled ones, which proves the validity of our entanglement-driving approach. It's also clear that the MO-12,13 and MO-15,16 are bonding-antibonding pairs, respectively, since there are obvious orbital entanglement within each orbital pair as shown in Fig.~\ref{CuCl2_entanglement}.

\begin{table}[htb]
    \caption{Important determinants of states $3^{2}\Sigma^{+}_{g}$ (a) and $4^{2}\Sigma^{+}_{g}$ (b) of CuCl$_{2}$ molecule, with orbitals grouped by irreducible representations of $D_{2h}$ symmetry. }
    \begin{tabular}{ccc}
        \hline
        \hline
        Determinant & Coefficient ($3^{2}\Sigma^{+}_{g}$) & Coefficient ($4^{2}\Sigma^{+}_{g}$) \\
        \hline
        $\rm |222d00,2,2,20,2,220,200\rangle$ &  0.541808       &  0.534725       \\
        $\rm |222d00,2,2,20,2,200,220\rangle$ & -0.538178       &  0.538394       \\
        $\rm |222d00,2,2,20,2,220,du0\rangle$ & -0.230630       & -0.222945       \\
        $\rm |222d00,2,2,20,2,du0,220\rangle$ &  0.229098       & -0.224485       \\
        $\rm |222d00,2,2,20,2,220,020\rangle$ & -0.227210       & -0.241580       \\
        $\rm |222d00,2,2,20,2,020,220\rangle$ &  0.225559       & -0.243068       \\
        \hline
        \hline
    \end{tabular}
    \label{CuCl2_excitation_patterns}
\end{table}

\begin{figure}[htb]
    \begin{center}
        \includegraphics[width=\textwidth]{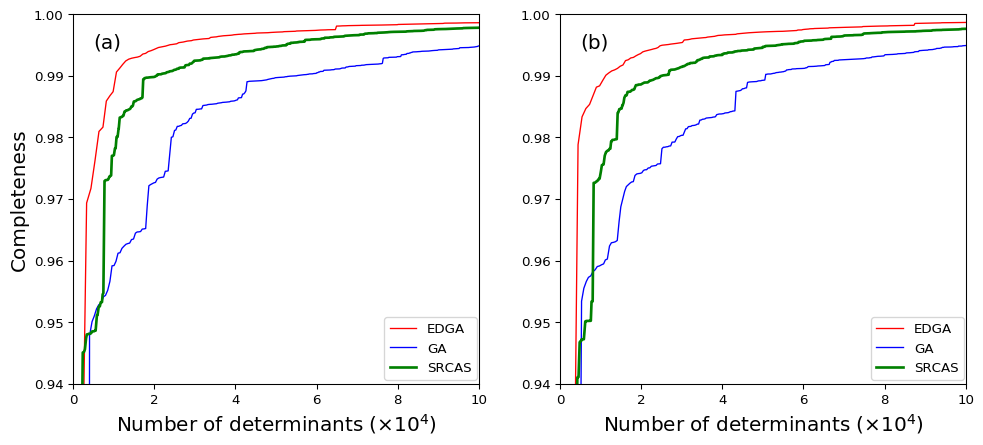}
    \end{center}
    \caption{Comparison of completeness of original GA, EDGA and SRCAS sampling schemes respecting to the number of determinants with coefficients larger than $1e^{-6}$ collected in the sampling routine. The EDGA scheme uses entanglement entropy data of canonical orbitals from DMRG(21,17)[1000]-SCF calculations on the states $3^{2}\Sigma^{+}_{g}$ (a) and $4^{2}\Sigma^{+}_{g}$ (b) of CuCl$_{2}$ molecule.}
    \label{CuCl2}
\end{figure}

When reconstructing the determinants, we also kept an eye on the efficiency. 
The relationship between the completeness and the number of determinants collected in the searching process is displayed in Fig.~\ref{CuCl2} for $3^{2}\Sigma^{+}_{g}$ and $4^{2}\Sigma^{+}_{g}$ states. 
It can be found that the SRCAS procedure already shows a very good performance in reconstructing the determinants. For example, around 20,000 determinants can reached 99\% completeness for the $3^{2}\Sigma^{+}_{g}$ state and 30,000 determinants for $4^{2}\Sigma^{+}_{g}$ state.
Furthermore, our EDGA scheme can reach the same completeness by collecting about 10,000 determinants less than the SRCAS scheme. The EDGA scheme can easily reach 99.5\% completeness by collecting about 30,000 determinants to obtain a very good approximation for the total wavefunction for both the two states.

\subsection{Eu-BTBP(NO$_3$)$_3$ complex}
In this section, we turn to the europium complex Eu-BTBP(NO$_3$)$_3$, in which the BTBP is one of the popular ligands for electively extract trivalent actinides (An) over lanthanide (Ln) fission products by solvent extraction via nitric acid solutions to organic solvents.
Nowadays, BTBP has been considered as one of the most promising species for partitioning the minor actinides from radioactive waste \cite{Hancock2012The, Panak2013Complexation, Jones2012ChemInform}. 
However, we are focus on its electronic structure rather than selectively in this work.

\begin{figure}[htb]
    \begin{center}
        \includegraphics[scale=1.00,bb= 0 0 500 255]{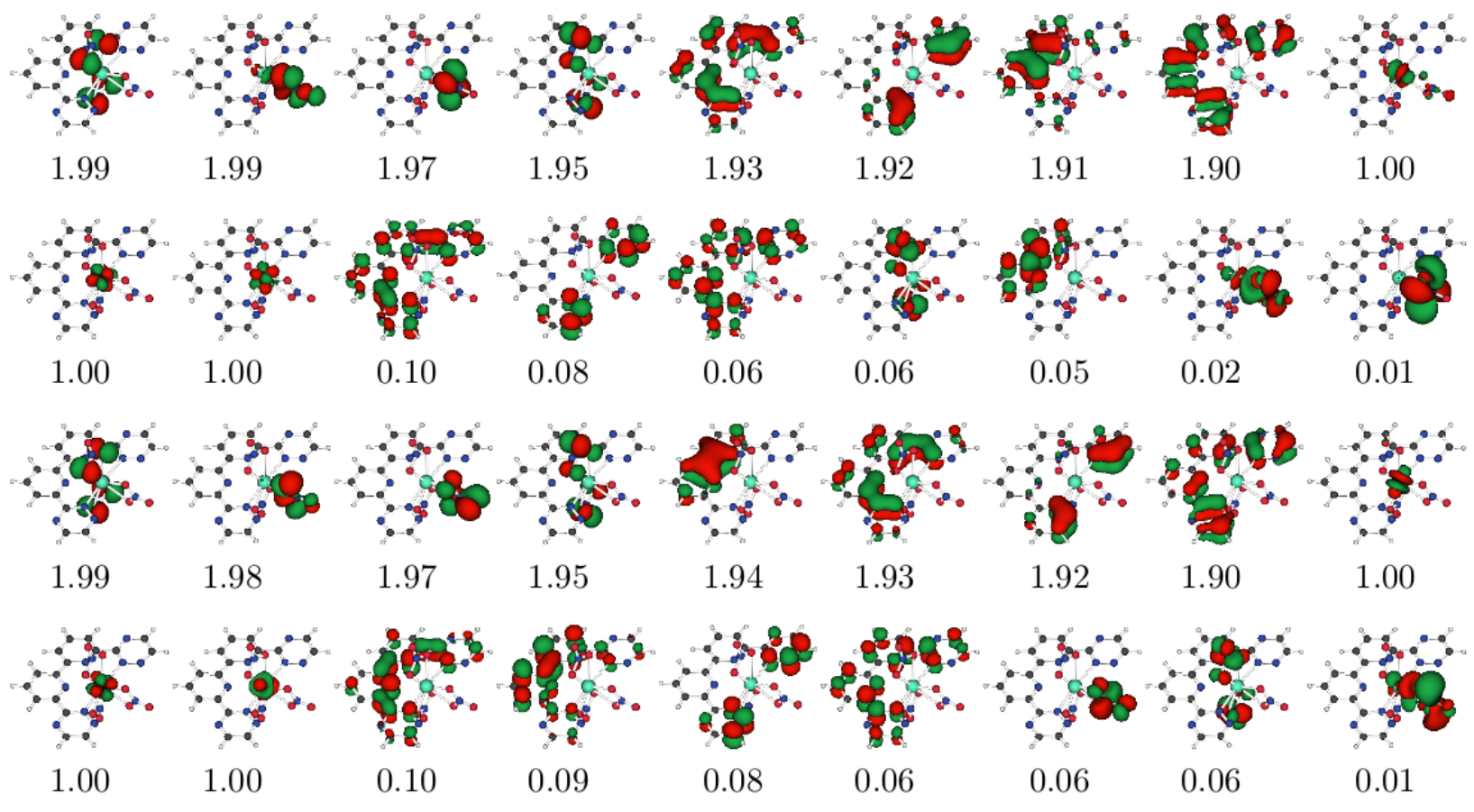}      
    \end{center}    
    \vspace{-1.0cm}
    \caption{The optimized NOs and NOONs from DMRG(38,36)[1000]-SCF calculation for $^7B_1$ state of Eu-BTBP(NO$_3$)$_3$. The first and second rows are irreps-\textit{a} orbitals, the third and forth rows are irreps-\textit{b} orbitals in $C_{2}$ symmetry. Similar optimized orbitals can be observed for other states.}
    \label{euorbDMRG}
\end{figure}

\begin{figure}[!htb]
    \begin{center}
        \includegraphics[scale=0.65,bb= 0 0 750 750]{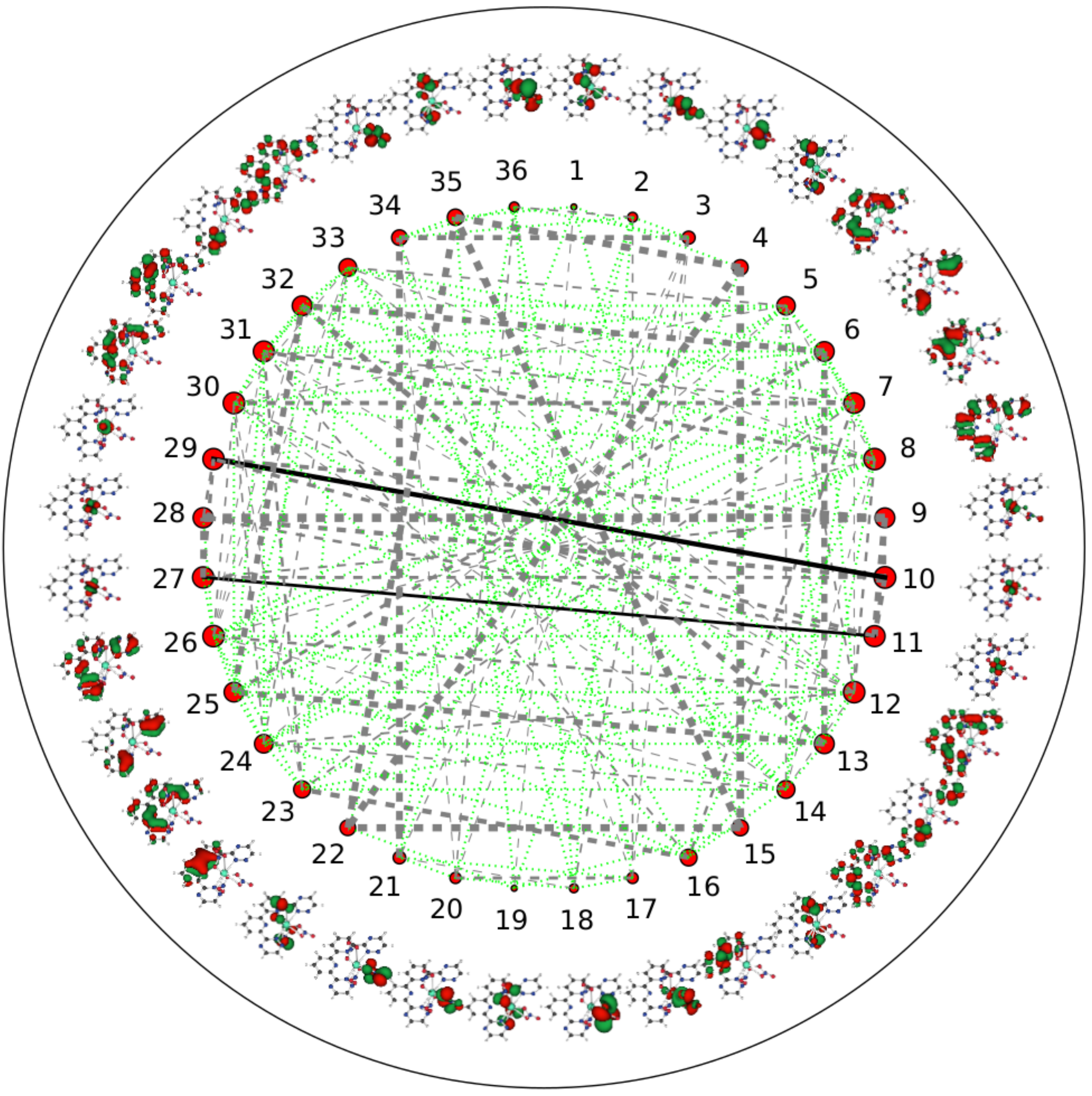}  
    \end{center}    
    \vspace{-1.0cm}
    \caption{The entanglement of $^7B_1$ state of Eu-BTBP(NO$_3$)$_3$ from DMRG(38,36)[1000]-SCF calculation. The 1-18 orbitals are irreps-\textit{a} orbitals, 19-36 orbitals are irreps-\textit{b} orbitals in $C_{2}$ symmetry. The size of the red circles indicates the magnitude of the values of single-orbital entropy $S^{(1)}_{i}$, and the color and thickness of the lines indicate mutual information $I_{ij}$.}
    \label{euorbENT}
\end{figure}

As the start, DMRG(38,36)[1000]-SCF calculations were employed with carefully selected orbitals in order to decide the ground state. The calculated electronic energies can be found in Tab.~\ref{tabEU}. It can clearly find that the maximum spin multiplicity state $^7B_1$ is the lowest energy electronic state, as reported by Narbutt and Oziminski \cite{narbutt2012selectivity}. Using the EDGA to reconstruct the determinants, we can have a more detailed understanding of this electronic state. The 10 most contributed determinants are listed in Tab.~\ref{BTBP_Patterns}, the optimized corresponding orbitals are illustrated in Fig.~\ref{euorbDMRG}, and mutual information and single orbital entropies $S^{(1)}_{i}$ are illustrated in Fig.~\ref{euorbENT}. It is clear that the wavefunction has obvious multireference character and these 10 determinants cover only 68.9\% of the whole CI expansion space, while about 9000 determinants cover 90\%.
By distinguishing the entangled orbital pairs in Fig.~\ref{euorbENT} and referring to the determinants in Tab.~\ref{BTBP_Patterns}, we can find that the electron excitations with large contribution do occur in entangled pairs. For example, the second determinant in Tab.~\ref{BTBP_Patterns} implies two electrons in MO-21 transfer to MO-34, and both the two MOs belong to the NO$_{3}$ ligand.

\begin{table}[!hbp]\centering 
    \caption{The calculated electronic energies (in Hartree) of Eu-BTBP(NO$_3$)$_3$ for different states with various spin multiplicities and irreducible representations by DMRG(38,36)[1000]-SCF calculation. All the energy results should be subtracted by 12710.00 Hartree.}
    \begin{tabular}{ccccccc}
        \hline
        \hline   
        State && $^1A_1$  & $^3A_1$ &  $^5A_1$ &  $^7A_1$  \\ 
        \cline{1-6}
        Energy && -9.968373 & -10.0269954    & -10.026785  &  -9.974740     \\   
        \hline
        \hline   
        State &&  $^1B_1$ &  $^3B_1$  & $^5B_1$ &  $^7B_1$  \\ 
        \cline{1-6}
        Energy && -9.941314   & -9.9411189   &  -10.023320    &   -10.113004   \\   
        \hline
        \hline   
    \end{tabular}
    \label{tabEU}
\end{table}

\begin{table}[htb]
    \caption{Important determinants of $^7B_1$ state of Eu-BTBP(NO$_3$)$_3$ complex from DMRG(38,36)[1000]-SCF calculation, with orbitals grouped by irreducible representations of $C_{2}$ symmetry. }
    \begin{tabular}{ccc}
        \hline
        \hline
        Determinant                           & Coefficient \\
        \hline
        $\rm |22222222uuu0000000,22222222uuu0000000\rangle$ & -0.8057339  \\
        $\rm |22222222uuu0000000,22022222uuu0000200\rangle$ &  0.0895062  \\
        $\rm |22202222uuu0000000,22222222uuu0000020\rangle$ &  0.0641611  \\
        $\rm |222d2222uuu000u000,222u2222uuu00000d0\rangle$ & -0.0640661  \\
        $\rm |222u2222uuu000d000,222d2222uuu00000u0\rangle$ & -0.0640522  \\
        $\rm |222d2222uuu000d000,222u2222uuu00000u0\rangle$ &  0.0633802  \\
        $\rm |22022222uuu0000000,22222222uuu0000200\rangle$ &  0.0629711  \\
        $\rm |22222222uuu0002000,22202222uuu0000000\rangle$ &  0.0626438  \\
        $\rm |22222222uuu0000000,22222202uuu0020000\rangle$ &  0.0600976  \\
        $\rm |22222u22uuu0d00000,222222d2uuu00u0000\rangle$ & -0.0599740  \\
        \hline
        \hline
    \end{tabular}
    \label{BTBP_Patterns}
\end{table}

\section{Conclusion}
In order to improve the sampling efficiency of reconstructing CI expansion for DMRG-MPS wave functions with a large active space, in this work we propose a new methodology of EDGA by virtue of combining the features of the “mutation” and “crossover” in gene algorithm and the concepts of entanglement in quantum information theory.

Our EDGA test calculations for the ground and excited states of various conjugated or transition metal compounds (such as polyacetylene, heptacene and CuCl2) verifies that it is feasible for our EDGA to reach a very high completeness (99\% or even higher) for CI expansion but only sampling a very small portion (e.g. $1/10^{12}$ for Heptacene) of the total Hilbert space. Comparison with traditional SRCAS also illustrated the remarkable improvements for efficiency and stability in our EDGA calculations for large active spaces.  

Therefore, EDGA can be expected to be a useful analysis tool for DMRG wave functions with a very large number of active orbitals.

\section{Acknowledgement}
The work was supported by the National Natural Science Foundation of China (Grant Nos. 21373109 and 21673109).

\bibliography{EDGA}
\end{document}